\documentclass{article}
\usepackage{spconf,amsmath,amssymb,graphicx,booktabs,hyperref,float,placeins}
\usepackage{cite,microtype}
\usepackage{siunitx,multirow,balance}
\graphicspath{{figures/}}
\hypersetup{hidelinks,hypertexnames=false}
\newcommand{\Hnull}{\mathcal H_0}
\newcommand{\Halt}{\mathcal H_1}
\newcommand{\dd}{\,\mathrm d}

\title{DATA-DRIVEN LIKELIHOOD-RATIO CONSTRUCTION FOR RADAR DETECTION\\
VIA BERNSTEIN INDUCTION OPERATORS}
\name{Yuxin Xu, Jun Tang, Ting Bai\thanks{This work has been submitted to the IEEE for possible publication. Copyright may be transferred without notice, after which this version may no longer be accessible.}}
\address{Department of Electronic Engineering, Tsinghua University\\
Beijing 100084, China}

\begin{document}
\ninept
\maketitle

\begin{abstract}
Accurate likelihood-ratio (LR) construction is challenging when clutter statistics are unknown. Prespecified clutter models may not adequately describe the observed data, and evaluating target-present densities may require computationally expensive marginalization. This paper proposes a data-driven LR construction framework based on a Bernstein polynomial model (BPM) and a Bernstein induction operator (BIO). The BPM weights are estimated from clutter-only samples, whereas the target-induced responses of the Bernstein basis densities are approximated offline in the same basis to construct column-stochastic matrices. These matrices map clutter-density weights to target-present density weights at specified signal-to-clutter ratios (SCRs), and mixtures of the matrices account for target fluctuations. For a fixed target model and basis representation, the matrix library is independent of the clutter-density weights and can be reused across clutter distributions. The resulting LR construction avoids repeated online evaluation of the physical-induction integrals. Simulations under analytical clutter models show close agreement with the reference LRs and detection probabilities, with lower online computation times in the non-Gaussian cases considered. Experiments using synthetic target echoes embedded in measured sea clutter further show improved detection probabilities relative to the evaluated baselines in the weak-target regime.
\end{abstract}


\section{Introduction}
\label{sec:introduction}
Radar target detection is commonly formulated as a binary hypothesis test that distinguishes clutter-only observations from observations containing both clutter and a target echo. Common models for clutter amplitudes include the Rayleigh, K, Weibull, lognormal, and $\mathcal{G}^{0}$ distributions~\cite{jakeman1976nonrayleigh,frery1997g0,angelliaume2019modeling}. Model-based construction of the target-present density requires models for both clutter and target echoes and often involves marginalization over the target phase and amplitude fluctuations. Two challenges arise: a prespecified distribution may not adequately describe the clutter, and the associated integrals may make repeated likelihood evaluations computationally expensive at different signal-to-clutter ratios (SCRs). These challenges motivate density representations that support both accurate clutter modeling and efficient likelihood-ratio (LR) computation.

Likelihoods derived from specific clutter models are widely used in radar detection. Conte et al. investigated the detection of signals with unknown parameters in K-distributed clutter~\cite{conte1991kglrt}. Yi et al. derived LRs for $\mathcal{G}^{0}$ clutter with Swerling~0/I targets and developed computationally efficient approximations for track-before-detect processing~\cite{yi2017g0tbd}. These likelihood formulations and their evaluation procedures depend on the assumed clutter and target models.

Nonparametric methods, including KDE, histograms, and log\-spline estimators, provide flexible density representations without prescribing a parametric clutter family~\cite{izenman1991nonparametric,scott1979histograms,kooperberg1991logspline}. For radar applications, Lang et al. used adaptive composite cubic B\'ezier curves to model sea-surface return distributions in synthetic aperture radar imagery~\cite{lang2016bezier}. Bai and Tang combined the monitoring of background changes with online KDE and detector calibration for nonstationary clutter~\cite{baiDataDrivenOnlineFramework2026}. Although these approaches provide flexible clutter models, estimation of the target-present density additionally requires target-present samples or an explicit model of the target contribution.

Alternatively, direct density-ratio methods ~\cite{sugiyama2008kliep,kanamori2009ulsif} avoid separately recovering both densities. Calibrated discriminative classifiers can also approximate likelihood ratios using samples generated by a simulator~\cite{cranmer2015ratios}. If a separate estimator is trained for each target condition, changes in target strength require additional target-present samples and repeated ratio estimation, with variability arising from simulation and fitting.

This paper develops a framework that combines data-driven clutter-density estimation with a reusable representation of physical induction. The main contributions are summarized as follows:
\par\noindent (1) We use a Bernstein polynomial model (BPM) to estimate the clutter density from clutter-only samples without assuming a specific parametric clutter distribution.
\par\noindent (2) We construct a Bernstein induction operator (BIO) that maps clutter-density weights to target-present density weights through a column-stochastic matrix. For a fixed target model and basis representation, the resulting matrix library can be reused across clutter distributions.
\par\noindent (3) We derive an explicit LR expression in the shared Bernstein basis, enabling online evaluation without repeated computation of the physical-induction integrals.

\begin{figure*}[!t]
 \centering
 \includegraphics[width=0.95\textwidth,trim=2pt 2pt 2pt 2pt,clip]{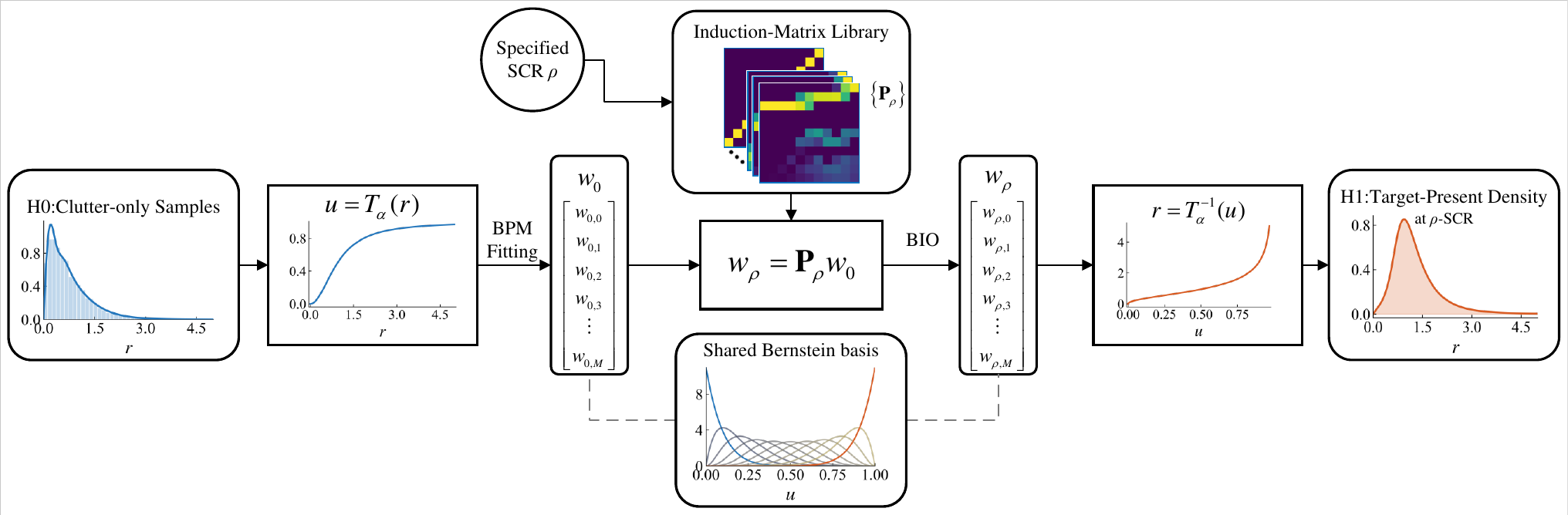}
 \caption{BIO framework for constructing target-present amplitude densities from clutter-only samples.}
 \label{fig:conditional-density-framework}
\end{figure*}

\section{Problem Formulation}
\label{sec:problem-model}
Let $C$ denote a circularly symmetric complex clutter sample normalized such that $\mathbb E|C|^2=1$. The received sample $Z$ is modeled under the two hypotheses as
\begin{equation}
 \begin{aligned}
  \Hnull &: Z=C,\\
  \Halt &: Z=C+\sqrt{\rho}\,e^{\mathrm j\Phi},
  \qquad \Phi\sim\operatorname{Unif}[0,2\pi),
 \end{aligned}
 \label{eq:observation-model}
\end{equation}
where $\Phi$ is independent of $C$ and $\rho\geq0$ denotes the SCR on a linear scale. Under the unit-power normalization, the target amplitude is $\sqrt\rho$. Let $R_0=|C|$ denote the clutter amplitude, with density $p(r\mid0)$. Under $\Halt$, the amplitude $R_\rho=|Z|$ has density $p(r\mid\rho)$. With $d_\rho(r,\theta)$ defined as $|r e^{\mathrm j\theta}-\sqrt\rho|$, circular symmetry yields
\begin{equation}
 p(r\mid\rho)
 =\frac{r}{2\pi}\int_0^{2\pi}
 \frac{p\!\left(d_\rho(r,\theta)\mid0\right)}
 {d_\rho(r,\theta)}\dd\theta,
 \qquad r\geq0.
 \label{eq:conditional-density-integral}
\end{equation}
Equation~\eqref{eq:conditional-density-integral} defines a linear physical-induction operator $\mathcal A_\rho$ that maps the clutter-amplitude density to the target-present amplitude density, i.e., $p(\cdot\mid\rho)=\mathcal A_\rho p(\cdot\mid0)$.

Target-power fluctuations are modeled by mixing the fixed-SCR densities according to a Swerling model:
\begin{equation}
 p(r\mid\bar\rho,\mathsf S)
 =\int_0^\infty p(r\mid\gamma)\,
 \pi_{\mathsf S}(d\gamma\mid\bar\rho).
 \label{eq:fluctuating-density-mixture}
\end{equation}
Here, $\bar\rho$ denotes the mean SCR, and $\pi_{\mathsf S}(d\gamma\mid\bar\rho)$ is the probability distribution of the instantaneous SCR $\gamma$ under fluctuation model $\mathsf S$.

\section{Method}
\label{sec:implementation}
\subsection{Bernstein Clutter Representation}
\label{subsec:bernstein-density}
The BPM represents a density as a convex combination of $M+1$ beta basis densities~\cite{vitale1975bernstein}. To represent amplitudes on $[0,\infty)$, we map them to $[0,1)$ using a hyperbolic transformation:
\begin{equation}
 \begin{aligned}
 \beta_{j,M}(u)&=(M+1)\binom{M}{j}u^j(1-u)^{M-j},\\
 u=T_\alpha(r)&=\tanh^2\!\left(\alpha\operatorname{asinh}r\right).
 \end{aligned}
 \label{eq:bernstein-beta-basis}
\end{equation}
Here, $j=0,\ldots,M$, $r\geq0$, $u\in[0,1)$, and $\alpha>0$ controls the compression of the amplitude axis. The basis densities and their mixture in the original amplitude domain are given by
\begin{equation}
 \begin{aligned}
 \psi_{j,M}(r)&=\beta_{j,M}\!\left(T_\alpha(r)\right)T_\alpha'(r),\\
 p_{\boldsymbol w}(r)&=\boldsymbol\psi_M(r)^{\mathsf T}\boldsymbol w,
 \qquad \boldsymbol w\in\Delta_M,
 \end{aligned}
 \label{eq:bernstein-density}
\end{equation}
where $\boldsymbol\psi_M=[\psi_{0,M},\ldots,\psi_{M,M}]^{\mathsf T}$ and $\Delta_M=\{\boldsymbol w:\boldsymbol w\succeq0,\mathbf1^{\mathsf T}\boldsymbol w=1\}$ is the probability simplex. For clutter-only samples $\{r_i\}_{i=1}^N$, set $u_i=T_\alpha(r_i)$. With the basis vector $\boldsymbol\beta_M(u)=[\beta_{0,M}(u),\ldots,\beta_{M,M}(u)]^{\mathsf T}$, maximum likelihood gives
\begin{equation}
 \widehat{\boldsymbol w}_0
 =\arg\max_{\boldsymbol w\in\Delta_M}
 \sum_{i=1}^{N}\log\!\left[
 \boldsymbol\beta_M(u_i)^{\mathsf T}\boldsymbol w\right].
 \label{eq:h0-weight-mle}
\end{equation}
For a fixed transformation, the weight-independent Jacobian is omitted. The concave problem can be solved by expectation-maximization (EM) or Newton-type methods~\cite{guan2016bernstein}, yielding $\widehat p(r\mid0)=\boldsymbol\psi_M(r)^{\mathsf T}\widehat{\boldsymbol w}_0$.

\subsection{Bernstein Induction Operator}
\label{subsec:alternative-density-generation}
\label{subsec:induction-operator}
By the linearity of $\mathcal A_\rho$, the induced mixture density is the weighted sum of the induced basis densities. The BIO approximates each induced basis density in the Bernstein basis of degree $M$ and assembles the resulting coefficient vectors as the columns of a matrix:
\begin{equation}
 \begin{aligned}
 \relax[\mathcal A_\rho\widehat p(\cdot\mid0)](r)
 &=\sum_{j=0}^{M}\widehat w_{0j}
   (\mathcal A_\rho\psi_{j,M})(r),\\
 (\mathcal A_\rho\psi_{j,M})(r)
 &\approx\boldsymbol\psi_M(r)^{\mathsf T}\boldsymbol p_{\rho,j},\\
 \mathbf P_\rho&=[\boldsymbol p_{\rho,0},\ldots,\boldsymbol p_{\rho,M}],\\
 \widehat{\boldsymbol w}_\rho&=\mathbf P_\rho\widehat{\boldsymbol w}_0.
 \end{aligned}
 \label{eq:online-induction}
\end{equation}
Since $\boldsymbol p_{\rho,j}\in\Delta_M$, $\mathbf P_\rho$ is column-stochastic, i.e., $\mathbf P_\rho\succeq0$ and $\mathbf1^{\mathsf T}\mathbf P_\rho=\mathbf1^{\mathsf T}$. The resulting approximation is $\widehat p(r\mid\rho)=\boldsymbol\psi_M(r)^{\mathsf T}\widehat{\boldsymbol w}_\rho$, which is nonnegative and integrates to one. Substituting this representation into~\eqref{eq:fluctuating-density-mixture} yields
\begin{equation}
 \begin{aligned}
 \overline{\mathbf P}_{\bar\rho,\mathsf S}
 &=\int_0^\infty\mathbf P_\gamma\,
   \pi_{\mathsf S}(d\gamma\mid\bar\rho),\\
 \widehat p(r\mid\bar\rho,\mathsf S)
 &=\boldsymbol\psi_M(r)^{\mathsf T}
   \overline{\mathbf P}_{\bar\rho,\mathsf S}\widehat{\boldsymbol w}_0.
 \end{aligned}
 \label{eq:fluctuating-conditional-density}
\end{equation}
Under unit-power normalization, all basis densities are subjected to the same target amplitude $\sqrt\rho$; hence, for fixed $(\alpha,M,\rho)$ and target model, $\mathbf P_\rho$ is independent of the clutter weights and can be precomputed for reuse across clutter distributions.

To construct the $j$th column of $\mathbf P_\rho$, we approximate the corresponding induced density in the transformed domain by maximizing the expected log-likelihood:
\begin{equation}
 \begin{aligned}
 \boldsymbol p_{\rho,j}^{\star}
 &=\arg\max_{\boldsymbol v\in\Delta_M}
 \frac{1}{2\pi}\int_0^1\!\int_0^{2\pi}\beta_{j,M}(u)\\[-0.2em]
 &\quad\times\log\!\left[
 \sum_{k=0}^{M}v_k\beta_{k,M}\!\left(V_\rho(u,\theta)\right)
 \right]\dd\theta\dd u .
 \end{aligned}
 \label{eq:operator-column-population}
\end{equation}
The transformed target-present amplitude is given by
\[
\begin{aligned}
V_\rho(u,\theta)&=T_\alpha(Y_\rho(u,\theta)),\\
Y_\rho(u,\theta)&=\sqrt{R(u)^2+\rho+2R(u)\sqrt\rho\cos\theta},
\end{aligned}
\]
where $R(u)=T_\alpha^{-1}(u)$. The integral in~\eqref{eq:operator-column-population} is evaluated offline using deterministic numerical quadrature. Solving the resulting optimization problem for $j=0,\ldots,M$ and assembling the coefficient vectors yields $\mathbf P_\rho$.

\begin{figure}[t]
 \centering
 \includegraphics[width=0.99\columnwidth]{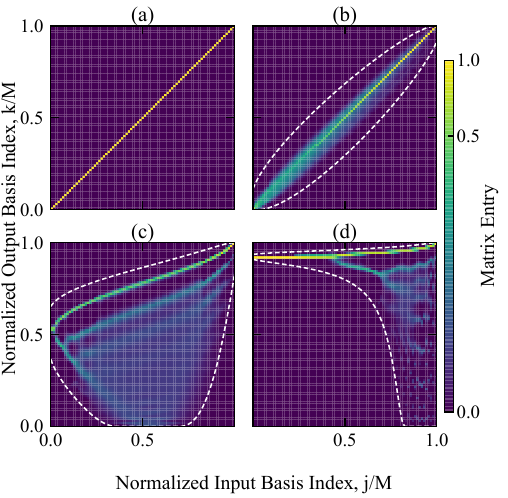}
 \caption{BIO matrices for $M=80$ and $\alpha=1.05$: (a) $\rho=0$;
 (b) $-10$~dB; (c) $0$~dB; and (d) $10$~dB. Dashed curves indicate the reference envelopes for the amplitude transformation.}
 \label{fig:operator-pattern-examples}
\end{figure}

Fig.~\ref{fig:operator-pattern-examples} shows representative BIO matrices at different SCRs. At $\rho=0$, we set $\mathbf P_0=\mathbf I$ to preserve the identity mapping exactly. As the SCR increases, the coefficient patterns change, with more weight generally assigned to larger output basis indices.

\subsection{Likelihood-Ratio Construction}
\label{subsec:likelihood-ratio-construction}
Given the estimated clutter-density weights, the approximate log-likelihood ratio (LLR) is
\begin{equation}
 \begin{aligned}
  \widehat\ell(r\mid\bar\rho,\mathsf S)
  &=\log\frac{\widehat p(r\mid\bar\rho,\mathsf S)}{\widehat p(r\mid0)}\\
  &=\log\frac{
  \boldsymbol\psi_M(r)^{\mathsf T}
  \overline{\mathbf P}_{\bar\rho,\mathsf S}
  \widehat{\boldsymbol w}_0}
  {\boldsymbol\psi_M(r)^{\mathsf T}
  \widehat{\boldsymbol w}_0}.
 \end{aligned}
 \label{eq:single-frame-operator-llr}
\end{equation}
The mixed operator $\overline{\mathbf P}_{\bar\rho,\mathsf S}$ is defined in~\eqref{eq:fluctuating-conditional-density}. For the nonfluctuating Swerling~0 (SW0) model, it reduces to $\mathbf P_{\bar\rho}$.

\section{Experiments}
\label{sec:experiments-applications}
\subsection{FSS Detection under Analytical Clutter Models}
\begin{figure}[!t]
 \centering
 \includegraphics[width=0.95\columnwidth]{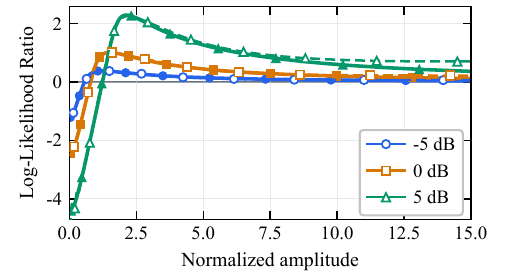}
 \caption{Single-scan LLRs in compound lognormal clutter at different SCRs. Solid lines: oracle LLRs; dashed lines: BIO approximations.}
 \label{fig:analytic-likelihood-lognormal}
\end{figure}

We evaluate LLR accuracy and fixed-sample-size (FSS) detection in unit-power Rayleigh, K, $\mathcal{G}^{0}$, and compound lognormal (CG-LN) clutter. We compare BIO-LLR, oracle LLRs from the true conditional densities, squared-amplitude accumulation, and log-amplitude accumulation~\cite{green1958logarithmic,hansen1972logarithmic} for SW0 and Swerling~I (SW1) targets; SW1 powers fluctuate independently across scans. Fig.~\ref{fig:analytic-likelihood-lognormal} illustrates BIO's approximation of nonmonotonic oracle LLRs at several SCRs.

For $L$ conditionally independent scans under each hypothesis, FSS detection uses threshold $\eta$:
\begin{equation}
 \widehat\Lambda_L(\bar\rho,\mathsf S)
 =\sum_{t=1}^{L}\widehat\ell(R_t\mid\bar\rho,\mathsf S)
 \mathop{\gtrless}_{\Hnull}^{\Halt}\eta .
 \label{eq:fixed-frame-llr}
\end{equation}
\begin{figure}[!t]
 \centering
 \includegraphics[width=0.95\columnwidth]{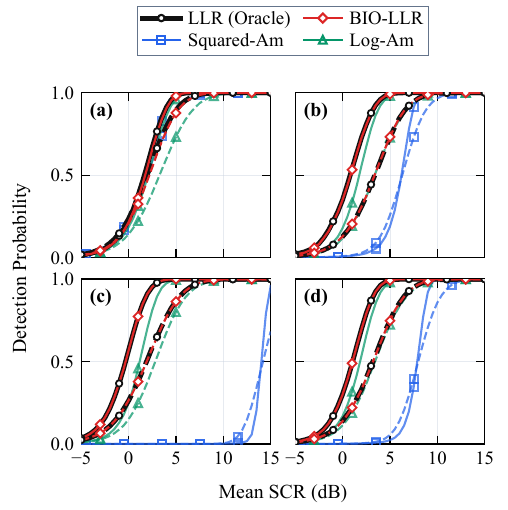}
 \caption{Detection probability versus mean SCR for FSS detection in (a) Rayleigh, (b) K, (c) $\mathcal{G}^{0}$,
 and (d) compound lognormal clutter. Solid lines: SW0 targets; dashed lines: SW1 targets.}
 \label{fig:sequence-detection-four-backgrounds}
\end{figure}

Both LLR detectors are matched to the tested SCR and fluctuation model. K and $\mathcal{G}^{0}$ texture shapes are $0.8$ and $1.5$, respectively; CG-LN texture has log-standard deviation $1$. We use $L=8$ and $\alpha=1.05$. Each background fit uses $5\,000$ clutter-only samples at $m_0=10$ to limit estimation complexity, followed by exact elevation to $M=80$~\cite{guan2016bernstein}, preserving the density while providing finer induction resolution. Thresholds use $50\,000$ independent $\Hnull$ calibration sequences at $P_{\mathrm{FA}}=10^{-3}$; each SCR uses $10\,000$ test sequences. Methods share samples, and results average 20 independent repetitions of background fitting and detection.

In Fig.~\ref{fig:sequence-detection-four-backgrounds}, squared-amplitude accumulation approaches oracle performance in Rayleigh clutter, whose LLR is monotonic. In the non-Gaussian cases, log-amplitude accumulation outperforms squared-amplitude accumulation, while BIO-LLR nearly coincides with the oracle under both target models.

Table~\ref{tab:llr-timing} compares median online LLR construction times over 20 independent background fits using one Intel Core Ultra~5~225H thread. Both methods cache clutter log-densities; BIO additionally caches the basis matrix and operator library. The model-specific baseline uses analytical expressions or numerical integration. BIO LLR accuracy is measured by weighted root-mean-square error (WRMSE) relative to the oracle, using an equal mixture of the two hypothesis densities, each normalized over the evaluation interval. BIO is faster in the tested non-Gaussian cases, with little runtime variation across backgrounds.

\begin{table}[!htbp]
\centering
\caption{Online LLR construction time (ms) and BIO WRMSE at 0~dB (1536 points).}
\label{tab:llr-timing}
\vspace{3pt}
\fontsize{9}{10}\selectfont
\sisetup{detect-weight=true,table-number-alignment=center}
\setlength{\tabcolsep}{1.9pt}
\renewcommand{\arraystretch}{0.95}
\begin{tabular}{@{}lcl S[table-format=1.3] S[table-format=1.3] S[table-format=2.3] S[table-format=2.3]@{}}
\toprule
 & & & {Rayleigh} & {$\mathcal{G}^{0}$} & {K} & {CG-LN} \\
\midrule
\multirow{4}{*}{\shortstack[l]{Runtime\\(ms)}} & \multirow{2}{*}{SW0} & BIO & \bfseries 0.036 & \bfseries 0.036 & \bfseries 0.036 & \bfseries 0.039 \\
 &  & Model-specific & 0.059 & 0.611 & 6.219 & 2.855 \\
\cmidrule(l){2-7}
 & \multirow{2}{*}{SW1} & BIO & 0.673 & \bfseries 0.660 & \bfseries 0.648 & \bfseries 0.708 \\
 &  & Model-specific & \bfseries 0.634 & 8.321 & 67.789 & 34.626 \\
\midrule
\multirow{2}{*}{\shortstack[l]{BIO\\WRMSE}} & SW0 & & 0.136 & 0.047 & 0.113 & 0.039 \\
 & SW1 & & 0.226 & 0.042 & 0.124 & 0.038 \\
\bottomrule
\end{tabular}
\end{table}

\subsection{Detection Using Measured Sea Clutter}
\begin{figure}[!t]
 \centering
 \includegraphics[width=0.95\columnwidth]{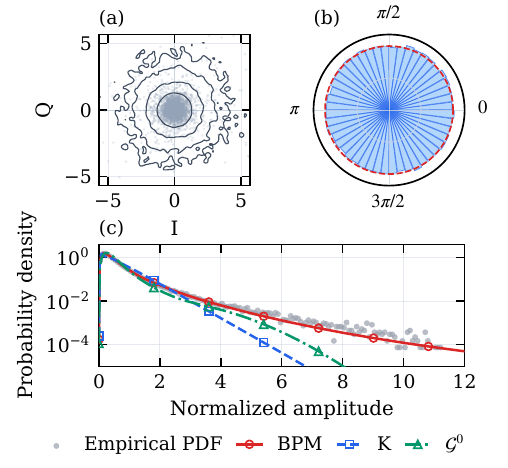}
 \caption{Measured clutter: (a) IQ density, (b) phase histogram, and (c) amplitude-density fits.}
 \label{fig:measured-h0-fit}
\end{figure}
To evaluate detection performance using measured clutter, we embed synthetic target echoes in measured sea-clutter data. We use 20 scans from the sea-clutter dataset described in~\cite{SeadetectingXbandRadar}. After local power normalization of the clutter, synthetic SW0 and SW1 target echoes are added to evaluate FSS detection with $L=8$. Scans 1--4, 5--12, and 13--20 are used for training, threshold calibration, and testing, respectively.

The BPM, K, and $\mathcal{G}^{0}$ clutter models are fitted using the same $50\,000$ training samples. The BPM is fitted at degree $m_0=20$ with $\alpha=1.05$, and the fitted density is represented exactly at the BIO degree $M=80$ by degree elevation. The in-phase and quadrature (IQ) density contours and phase histogram in Fig.~\ref{fig:measured-h0-fit} are consistent with approximate circular symmetry of the measured clutter. Over the displayed amplitude range, the BPM provides a closer fit to the empirical density in the tail than the fitted K and $\mathcal{G}^{0}$ models.

\begin{figure}[!htbp]
 \centering
 \includegraphics[width=0.95\columnwidth]{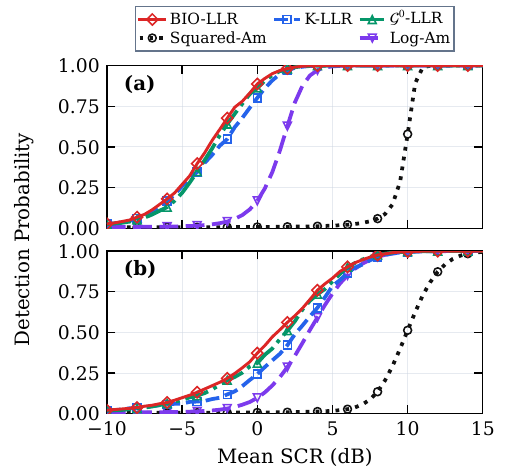}
 \caption{Measured-clutter detection with synthetic target echoes: (a) SW0; (b) SW1.}
 \label{fig:measured-pd-scr}
\end{figure}

A separate threshold is calibrated for each of the three LLR detectors and the two amplitude-based detectors using the held-out calibration set, with a nominal $P_{\mathrm{FA}}=0.005$. Fig.~\ref{fig:measured-pd-scr} shows that the three LLR detectors achieve higher detection probabilities than the amplitude-based detectors in the weak-target regime, with BIO-LLR attaining the highest values among the evaluated methods in this regime. For SW0 targets at an SCR of $-6$~dB, the detection probabilities of BIO-LLR, K-LLR, and $\mathcal{G}^{0}$-LLR are $0.180$, $0.167$, and $0.130$, respectively.

\section{Conclusion}
\label{sec:conclusion}
This paper has presented a BIO framework for data-driven construction of target-present densities and LRs. The clutter-density weights are estimated from clutter-only samples, and precomputed induction matrices map these weights to target-present density weights at specified SCRs. For a fixed basis representation and target model, the same matrix library can be reused across clutter distributions, avoiding repeated online evaluation of the physical-induction integrals. Simulations under analytical clutter models show detection probabilities close to those of the oracle detector and reduced online LLR construction times in the non-Gaussian cases considered. Experiments with synthetic target echoes embedded in measured sea clutter show higher detection probabilities than the evaluated baselines in the weak-target regime. Future work will investigate additional detection applications and more efficient construction of the induction matrices.

\balance

\end{document}